\documentclass[sigconf,natbib=true]{acmart}

\renewcommand\footnotetextcopyrightpermission[1]{}

\AtBeginDocument{%
}

\setcopyright{none}

\usepackage{algorithm}
\usepackage{algorithmic}

\usepackage{amsmath}

\usepackage{amssymb}

\usepackage{booktabs}
\usepackage{multirow}
\usepackage{graphicx}
\usepackage{subcaption}
\usepackage[most]{tcolorbox}
\usepackage{tabularx}
\usepackage{array}
\usepackage{xcolor}
\usepackage{colortbl}
\usepackage{lineno}
\usepackage{makecell}

\newcolumntype{Y}[1]{>{\centering\arraybackslash}p{#1}}
\newcolumntype{L}[1]{>{\raggedright\arraybackslash}p{#1}}
\newcolumntype{C}[1]{>{\centering\arraybackslash}p{#1}}
\newcolumntype{M}[1]{>{\centering\arraybackslash}m{#1}}

\renewcommand{\arraystretch}{0.55}

\begin{document}

\title{SAERec: Constructing Fine-grained Interpretable Intents Priors via Sparse Autoencoders for Recommendation}





\author{Jiangnan Xia}
\affiliation{
  \institution{University of Georgia}
  \country{United States}
}

\author{Xuansheng Wu}
\affiliation{
  \institution{Shanghai AI Laboratory}
  \country{China}
}

\author{Yu Yang}
\affiliation{
  \institution{The Education University of Hong Kong}
  \country{Hong Kong}
}

\author{Xin Wang}
\affiliation{
  \institution{Jilin University}
  \country{China}
}

\author{Ninghao Liu}
\affiliation{
  \institution{The Hong Kong Polytechnic University}
  \country{Hong Kong}
}

\renewcommand{\shortauthors}{Xia et al.}

\begin{abstract}
Intent-based recommender systems have gained significant attention for enhancing accuracy and interpretability by modeling the underlying motivations behind user behaviors. 
Most existing models derive intents directly from user sequences via clustering or prototype learning. 
However, they are sensitive to the sequences quality, require presetting the number of intents and lack explicit semantic grounding. 
These issues lead to an incomplete and coarse intent set and limit the effectiveness of recommendation. 
In this paper, we propose the Sparse Autoencoder for intent-based recommendation (\textbf{SAERec}), a novel recommender that automatically constructs a fine-grained, interpretable intent space from a textual corpus to guide recommendation. 
Rather than treating texts as side signals, SAERec leverages them as high-information-density evidence for intent construction. 
Specifically, we first extract a comprehensive set of fine-grained interpretable intents from the latent space of Large Language Models (LLMs) by using a Sparse Autoencoder (SAE) to disentangle and interpret the embedding of texts, which isolates intent-related semantics from substantial textual noise. 
Then, for each user, we retrieve relevant intents from the set as priors to guide recommendation. 
It contains personal intents matching user's current interests, and public intents capturing general item patterns shared across users (e.g., "quality", "price"). 
Finally, to integrate retrieved intents into sequence modeling, we propose a multi-branch attention mechanism that captures temporal dependencies and injects both personal and public intent signals, followed by an adaptive fusion layer to construct the final user representation for recommendation.  
Extensive experiments on public datasets demonstrate the superiority of SAERec, consistently outperforming state-of-the-art baselines while providing human-understandable explanations. 
Our code is available at \url{https://anonymous.4open.science/r/SAERec-CE84}.
\end{abstract}




\maketitle

\begin{figure}[!t]
  \centering
    \includegraphics[scale=0.42]{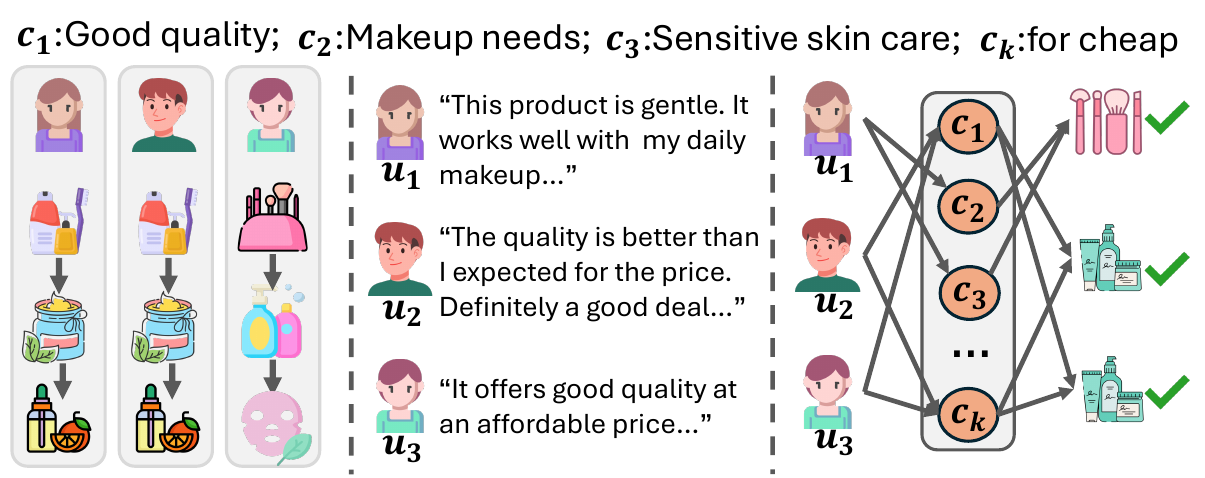}
    \caption{Illustration of intent-based recommendation, where $c_1, c_2, ..., c_k$ are underlying intents of users.}
  \label{fig:intro1}
\end{figure}

\section{Introduction}

Despite the success of recommenders in modeling user sequences \cite{zhao2024recommender, yin2025unleash}, user decisions in real-world scenarios are often influenced by \textit{underlying intents}, which plays a crucial role in shaping real-world user-item interactions \cite{wang2021learning}. 
For example, as shown in Figure~\ref{fig:intro1}, users with similar interaction patterns may arrive at different purchasing decisions when motivated by different intents (e.g., "Sensitive skin care" vs. "Cheap"), 
while users with different interaction histories can converge on the same item when sharing similar intent (e.g., "Cheap" and "Good quality"). 
User intents can be inferred from both historical interactions and textual feedback such as reviews. 
Compared to general recommendations, intent-based models introduce intents as an intermediate layer between user histories and item prediction, which have shown effectiveness in improving recommendation \cite{zhang2023efficiently,qin2024intent}.


Most studies directly infer intents from interactions, either by clustering user sequences or mapping them to a fixed set of learnable intent prototypes \cite{ma2020disentangled, li2023multi}. 
Despite capturing intents, these methods suffer from several limitations. 
First, the learned intents are tightly coupled with interaction quality. The sparse or noisy sequences will lead to unreliable intent representations, which degrade model performance \cite{jannach2024survey}. 
Second, they often require predefining the intent number, which significantly affects performance but is difficult to
determine in practice. 
An insufficient number may under-represent user diversity, while an overly large number often leads to redundant or noisy intent representations \cite{tan2021sparse, chen2022intent}. 
Third, the learned intents often lack
explicit semantics, hindering their interpretability and applicability in downstream scenarios \cite{zhang2023efficiently}. 
These issues could result in an \textit{incomplete and coarse intent set}. 
Recently, a few works have explored inferring user intents from textual data, which is semantically richer. 
Sun et al. \cite{sun2024large} employ large language models to extract interpretable intents from item titles in individual user sequences for re-ranking. 
However, intent inference is performed independently for each sequence, hindering cross-user semantic alignment and limiting the generalization of recommendations. 
And per-sequence LLM inference is latency- and cost-intensive. 
These limitations underscore the need for a new method to construct a high-quality intent set that is both scalable and generalizable.

To this end, we aim to automatically construct a comprehensive set of fine-grained interpretable intents from the textual corpus (e.g., user reviews) that inherently provide rich semantic clues about user preferences, and then use these intents to guide recommendations. 
While this route is promising, achieving it raises several challenges. 
First, a natural starting point is to encode the texts with a pretrained language model and treat the resulting embeddings as intent evidence. However, such embeddings are dense and entangled, mixing multiple user intents with contextual noise. This polysemantic nature makes it difficult to isolate distinct intent signals~\cite{arora2018linear}. 
Second, even if a comprehensive set of intents can be obtained, effectively leveraging them in sequential recommendations remains non-trivial. 
Intents are learned in a text-derived LLM latent space rather than the sequence model’s latent space, and the constructed intent set can be large (e.g., thousands of intents). 
Injecting intents indiscriminately risks cross-space mismatch and irrelevant signals that degrade personalization. 
It calls for explicit mechanisms to align spaces, select relevant intents per user and effectively inject them to sequence modeling for improved recommendation. 


To address these challenges, we propose the Sparse Autoencoder for intent-based recommendation (\textbf{SAERec}), a novel recommender that automatically constructs a comprehensive set of fine-grained interpretable intents and uses it to guide recommendations. Specifically, SAERec first leverages Sparse Autoencoders (SAE) \cite{huben2023sparse} to project dense text embeddings into a high-dimensional sparse space, where each latent dimension corresponds to a potentially disentangled vector. 
To interpret and extract intents without labels or a fixed number of intents, 
we associate each vector with its most informative text words from the task dataset, and then prompt LLM to label the semantic meaning of each vector based on these words. 
This enables automatic identification of useful intents based on human-understandable semantics. 
Then, we introduce a dual-level retrieval strategy that selects relevant intents for each user, which include personal intents that reflect the user’s specific interests and public intents that capture general patterns shared across users (e.g., "quality", "good value"). Together, these intents serve as \textit{informative priors} to improve both personalization and generalization in recommendation. 
Finally, to effectively inject the retrieved intent priors into sequence modeling, we design multi-branch attentions to learn user representations that align personal intents, public intents, and capture temporal dependencies. 
Experiments on multiple real-world benchmarks show that SAERec achieves superior accuracy and human-understandable explanations. 
The contributions of this work are summarized below: 
\begin{itemize}
    \item We propose SAERec, a novel intent-based recommender system. It automatically constructs a comprehensive set of fine-grained interpretable intents from textual corpus to guide recommendation. 
    \item We design an unsupervised pipeline that uses SAE to disentangle and interpret dense text embeddings of LLMs. Dedicated retrieval and injection mechanisms are proposed to enhance recommendation by selecting and integrating relevant intents with sequence learning. 
    \item We conduct extensive experiments, demonstrating that SAERec achieves strong recommendation accuracy and interpretable, human-aligned explanations. 
\end{itemize}

\section{Problem Statement}
The sequential recommendation problem is stated as below.
Let $\mathcal{U}$ and $\mathcal{V}$ denote the sets of users and items. For each user $u \in \mathcal{U}$, we observe a sequence of interactions $S_u = [v_1, v_2, \ldots, v_T]$, where $v_t \in \mathcal{V}$ is the interacted item at step $t$, and $T$ is the length of the sequence. The objective of sequential recommendation is to predict the next item $v_{T+1}$ that user $u$ is likely to engage, based on the user interaction history $S_u$.
In our work, we assume user reviews or item descriptions are available in the recommendation scenario, which is common in modern recommender systems.

\begin{figure*}[!t]
    \centering
    \includegraphics[scale=0.212]
    {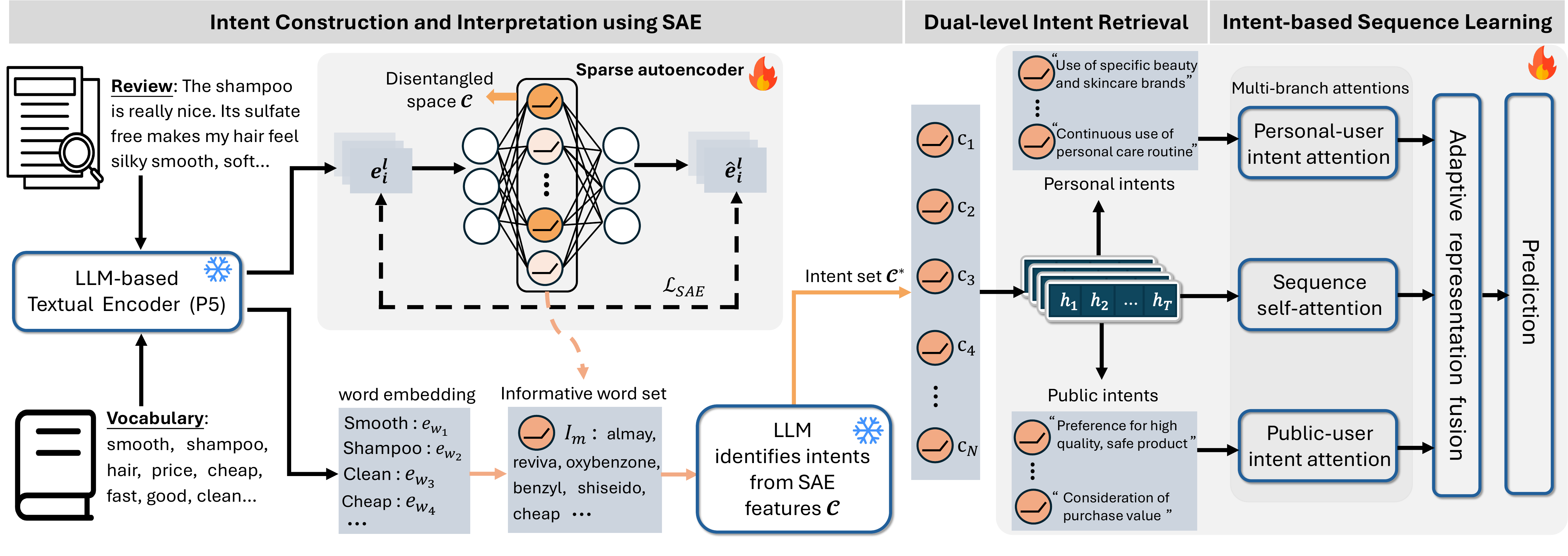}
    \caption{The overall pipeline of the novel Sparse Autoencoder for intent-based recommendation (SAERec). The proposed pipeline includes three stages: the fine-grained interpretable intent set construction,  dual-level intent retrieval, and intent-based sequence learning. 
    The intent set within reviews are automatically identified with SAE and LLMs, and then two types of relevant intents for each user are retrieved from the set. 
    Finally, the intents are integrated with sequence learning to generate recommendation results.}
    \label{framework}
\end{figure*}


\section{Methodology}
In this section, we introduce the proposed \textbf{SAERec} framework in detail. 
As shown in Figure~\ref{framework}, SAERec consists of three modules: fine-grained interpretable intent set construction, dual-level intent retrieval, and intent-based sequence learning. 
We make use of user reviews and item descriptions, and transform them into embeddings with an LLM-based encoder.. 
First, we train a Sparse Autoencoder (SAE) to disentangle textual embeddings, and construct the intent set automatically with the help of LLMs. 
Second, we introduce a dual-level intent retrieval strategy that selects user-specific (personal) intents and globally shared (public) intents for each user, supporting personalization and generalization, respectively. 
Finally, we design an injection mechanism that aligns the retrieved intents with sequence and adaptively fuses the resulting user representations to enhance both accuracy and explainability in recommendation. 
Notably, the intent set is constructed from the review corpus, instead of sequences, which is more comprehensive and stable and can still work if some users or items have no reviews.

\subsection{Intent-based Sequential Recommendation}
Intent-based recommendation is a fundamental module in our framework. 
It aims to capture the underlying motivations behind user behaviors as user intents~\cite{chen2022intent, qin2024intent}, and incorporates the intents into sequential modeling to improve recommendation quality.
Typically, in intent-based recommendation, a set of latent \textbf{intent vectors} $\mathcal{C}=\{\mathbf{c}_m\}_{m=1}^M$ is defined for the whole system, where each vector encodes a motivation or interest of users.
Then, the recommender system predicts the probability that a user interacts with an item $v_i$ as $P(v_i | S_u, \mathcal{C}_u)$, where $\mathcal{C}_u \subset \mathcal{C}$ is the set of activated intents influencing the user's behavior. 


\textbf{Challenges for Intent-based Modeling.} Despite the recent progress in intent-based recommendation~\cite{chen2022intent,liu2020explainable,ma2020disentangled}, several fundamental challenges remain.
(1)~Existing methods often rely on a small and fixed number of intents. This calls for a method that can construct \textbf{a comprehensive and fine-grained intent set} to represent diverse and nuanced user motivations.
(2)~Intents are often learned as latent clusters or prototype vectors whose meanings are opaque, which limits their use as reliable priors for downstream prediction. Addressing this issue requires learning interpretable intents with \textbf{clear semantic meanings}.
(3)~User-specific intents inferred directly from interactions could be noisy and unstable. Sparse or outdated user histories can also lead to unreliable intent estimation and hurt generalization. This calls for mechanisms that can balance personalized intent signals with more \textbf{stable, globally shared intent patterns}.
In this work, we address the above challenges in Section~\ref{subsec:sae}$\sim$ Section~\ref{subsec:dual}, respectively.

\subsection{Intents Construction with Sparse Autoencoders}\label{subsec:sae} 
In this part, our aim is to construct a comprehensive set of intents in an automatic and unsupervised manner. We utilize the rich semantic information encoded in LLMs and decompose their latent representations to extract meaningful intent signals.

\subsubsection{Encoding Intents with Text Embeddings.} 
We leverage review corpus and product descriptions as primary sources for intent extraction~\cite{chin2022datasets}. 
Compared to interaction sequences, textual data provide rich semantic evidence of user intent, e.g., users often explicitly express their feelings about items in reviews, making them suitable for the task. 
We first encode the texts using an LLM to obtain expressive semantic embeddings. 
However, general-purpose LLMs are not optimized for recommender systems~\cite{devlin2019bert, jiang2023mistral}, and may overlook decision-critical cues. 
Thus, we adopt recommendation-aware LLMs as textual encoders, such as P5 \cite{geng2022recommendation}, which are trained to preserve both semantic meaning and collaborative signals. 
Given the review corpus $\mathcal{R} = \{r_1,\dots, r_X\}$,
for each review $r_i$, its contextualized embedding is $\mathbf{e}_{i}^{l} = \text{P5}^l(r_i)\in\mathbb{R}^D$, where $D$ is the hidden dimension of representation and $\mathbf{e}_{i}^{l}$ is obtained by average-pooling the hidden states from a selected encoder layer $l$. 
Since P5 adopts an encoder-decoder architecture, we extract embeddings from the encoder, which has been shown to capture general and concept-level semantics \cite{tenney2019bert, liu2019linguistic}. 

\subsubsection{Extracting Intents via Disentangling Embeddings.} 
Although text embeddings encode rich intent-related information, they are inherently \textbf{polysemantic} \cite{ arora2018linear,scherlis2022polysemanticity}, as each dimension could entangle multiple latent intents and contextual noise. This semantic entanglement, along with the lack of labels, makes it difficult to isolate user intents.
Existing unsupervised approaches, such as sequence-level clustering~\cite{chen2022intent}, partially address this issue, but still suffer from entangled representations and are restricted to a small number of coarse intents.. 
This raises a key question: How can we extract a comprehensive set of disentangled intents from text embeddings without supervision? 

We address this by introducing a top-$K$ Sparse Autoencoder (SAE)
\cite{cunningham2023sparse}, which contains an encoder and a decoder, to project the LLM-derived text embeddings into a \textbf{high-dimensional feature space} with enforced sparsity. 
It imposes a semantic bottleneck that activates only the $K$ most salient features per input, pushing each features to specialize and encouraging the emergence of disentangled vectors, some of which potentially reflect intent-related signals. 
Formally, given an LLM-derived text embeddings $\mathbf{e}_i^l \in \mathbb{R}^D$,
the encoder produces a sparse feature vector $\mathbf{z}_i \in \mathbb{R}^M$, where $M$ denotes the number of features and $M \gg D$. 
Then, the decoder reconstructs $\mathbf{e}_i^l$ as $\hat{\mathbf{e}}_i^l$ based on $\mathbf{z}_i$. 
The auto-encoder and its training loss are defined as follows:
\begin{equation}
    \begin{split}
        \hat{\mathbf{e}}_i^l & = \mathbf{W}_{\text{dec}} \cdot \underbrace{\text{TopK}(\mathbf{W}_{\text{enc}} \mathbf{e}_i^l + \mathbf{b}_{\text{enc}})}_{\mathbf{z}_i} + \mathbf{b}_{\text{dec}}, \\
        \mathcal{L}_{\text{SAE}} & = \|\hat{\mathbf{e}}_i^l - \mathbf{e}_i^l\|_2^2,
    \end{split}
    \label{sae_train}
\end{equation}
where the encoder matrix is $\mathbf{W}_{\text{enc}} \in \mathbb{R}^{M \times D}$, the decoder matrix $\mathbf{W}_{\text{dec}}=\mathbf{W}_{\text{enc}}^{\top}$ uses shared weights, $\mathbf{b}_{\text{enc}}$ and $\mathbf{b}_{\text{dec}}$ are bias terms. The function TopK$(\cdot)$ zeros all but the top-$K$ activated features. All SAE parameters are optimized by minimizing $\mathcal{L}_{\text{SAE}}$.
After training, the encoder weight matrix $\mathbf{W}_{\text{enc}}$ implicitly defines a disentangled space $\mathcal{C} = \{ \mathbf{c}_1, \dots, \mathbf{c}_M \}$, where each \textbf{basis vector} $\mathbf{c}_m = \mathbf{W}^\intercal_{\text{enc}}[m, :]$ captures a certain semantic direction. The large $M$ provides overcompleteness, supplying enough atoms to allocate distinct dimensions to different semantic factors. 
These basis vectors provide a foundation for extracting user intents.

\subsection{Intent Filtering via Semantic Interpretation}\label{subsec:interpret}
The learned $\mathcal{C}=\{\mathbf{c}_m\}_{m=1}^{M}$ finds disentangled directions in the latent space of LLM embeddings.
Although these vectors capture distinct factors, their semantic meanings remain unknown. Many factors are \textit{noisy or irrelevant to recommendation tasks}, and cannot be used to represent user intents. 
To this end, we introduce an interpretation approach to understand the semantic meaning of each vector $\mathbf{c}_m$, and retain only those vectors whose meanings are relevant to the recommendation scenario. 
Specifically, we align $\mathbf{c}_m$ with a set of natural language words $\mathcal{I}_m \subset \mathcal{V}$, 
and optimize $\mathcal{I}_m$ via a mutual information-based objective~\cite{wu2025interpreting}:
\begin{equation}
\begin{split}
\max_{|\mathcal{I}_m| = Z} \text{MI}(\mathbf{c}_m;\mathcal{I}_m) & \propto \max_{|\mathcal{I}_m| = Z} \sum_{w \in \mathcal{I}_m} p(w | \mathbf{c}_m) \log p(\mathbf{c}_m |w), \\
\text{s.t.} \,\,\,\,\,\,\, p(w\mid \mathbf{c}_m) & =
\frac{\exp\!\big(\langle \mathbf{e}_w,\mathbf{c}_m\rangle\big)}
     {\sum_{w'\in\mathcal{V}}\exp\!\big(\langle \mathbf{e}_{w'},\mathbf{c}_m\rangle\big)}, \\
 p(\mathbf{c}_m\mid w) & =
\frac{\exp\!\big(\langle \mathbf{e}_w,\mathbf{c}_m\rangle\big)}
     {\sum_{m'=1}^{M}\exp\!\big(\langle \mathbf{e}_w,\mathbf{c}_{m'}\rangle\big)},
\end{split}
\end{equation}
where $\mathcal{V}$ denotes the vocabulary and $Z$ is the size of $\mathcal{I}_m$; 
$\mathbf{e}_w \in \mathbb{R}^D$ is the embedding of word $w$; $\langle \cdot, \cdot \rangle$ is the dot product. 
By intuition, $p(w|\mathbf{c}_m)$ measures how strongly the word $w$ is associated with the vector $\mathbf{c}_m$, and $p(\mathbf{c}_m|w)$ measures how uniquely this association identifies $\mathbf{c}_m$ among all vectors. 
Together, this objective favors the set of words $\mathcal{I}_m$ that are both strongly and uniquely aligned with a given vector.

We then identify which vectors correspond to meaningful user intents based on their textual explanations $\{\mathcal{I}_m\}_{m=1}^{M}$. 
Manually performing this identification is impractical, as the number of learned vectors $M$ typically ranges from tens of thousands to hundreds of thousands. 
To solve this problem, we employ a language model $g$ to automatically judge whether a given vector represents a meaningful user intent for recommendation according to a guideline written by human experts. 
This strategy is well motivated because we have provided concise and human-readable explanations $\mathcal{I}_m$ for each learned vector $\mathbf{c}_m$, and modern LLMs are capable of understanding and reasoning over such textual descriptions. 
Specifically, we first prompt the LLM to summarize the semantic meaning of each vector based on its aligned word set $\mathcal{I}_m$, and then ask the LLM to determine whether the summarized semantics correspond to a meaningful recommendation intent. 
Only vectors identified as meaningful intents are retained in the final intent set $\mathcal{C}^*$. 
More details about prompting and filtering criteria are provided in Appendix~\ref{ap_prompt}. 
Till now, we finish building the intent set $\mathcal{C}^*$.

\subsection{Dual-Level Intent Retrieval for Users}\label{subsec:dual}
Given the constructed intent set $\mathcal{C}^*$ and a user $u$, we are able to identify relevant intents based on the user's historical interaction $S_u = [v_1, \ldots, v_T]$, called \textbf{personal intent set} $\smash{\mathcal{C}_{per}^{(u)}\subset\mathcal{C}^*}$.
Besides, to enhance generalization, we will also retrieve a \textbf{public intent set} $\mathcal{C}_{pub}\subset\mathcal{C}^*$ that captures general behavior patterns shared across users. The two intent sets form a dual-level retrieval approach.

\subsubsection{Personal Intent Retrieval}
We aim to retrieve a set of intents that semantically aligns with the user’s current behavior. 
A na\"ive approach is to encode the user sequence with a language model such as P5, and then select the most activated latent vectors from $\mathcal{C}$. 
However, although P5 is a language model, here it operates on \textit{index-based sequences} (e.g., item IDs), lacking direct access to textual context. As a result, the sequence embeddings are not well aligned with intent vectors, leading to less effective retrieval results.
Thus, we align user sequence embeddings with the intent set $\mathcal{C}^*$ in a shared space. 
Specifically, we define a sequence encoder $f_\theta(\cdot)$, such as SASRec~\cite{kang2018self}, to encode the sequence $S_u$ as $\mathbf{H}_u = f_\theta(S_u) = [\mathbf{h}_1,...,\mathbf{h}_T]$. 
Meanwhile, we project each intent vector $\mathbf{c}_i\in\mathcal{C}^*$ into the same latent space via a learnable mapping $\mathbf{c}^\prime_i = \mathbf{W}_\text{map} \cdot \mathbf{c}_i$, where $\mathbf{W}_\text{map} \in \mathbb{R}^{D \times d}$ is the mapping matrix. We then compute a semantic alignment score between $\mathbf{h}_T$ and $\mathbf{c}^\prime_i$ via cosine similarity, and select the top-$S$ intents as the user's personal intent:
\begin{equation}
        \mathcal{C}_{per}^{(u)} = \text{TopS}_{\mathbf{c}_i \in \mathcal{C}^*} \{ \cos(\mathbf{h}_T, \mathbf{c}^\prime_i) \}.
    \label{per_se2}
\end{equation}
The resulting set $\mathcal{C}_{per}^{(u)}$ semantically aligned and personalized intent priors for downstream modeling.

\subsubsection{Public Intent Retrieval}
While personal intents capture fine-grained alignment, they are sensitive to noises and randomness in individual interaction sequences~\cite{jannach2024survey}, and may fail to capture generalizable semantics.
To address this, we further introduce a public intent set as a globally shared semantic prior. 
The public intents are selected based on their activation consistency across users. We estimate this consistency using an Exponential Moving Average (EMA) method~\cite{tarvainen2017mean}.
Compared to raw frequency counts or sliding-window statistics, EMA applies exponential decay to historical activations, emphasizing recent trends while retaining long-term patterns for stable relevance estimation. 
The retrieval is defined as follows: 
\begin{equation}
\mathcal{C}_{pub} = \text{TopS}_{\mathbf{c}_i \in \mathcal{C}^*} \left( \alpha \cdot \text{EMA}^{(b-1)}(\mathbf{c}_i) + (1 - \alpha)\cdot \text{Freq}^{(b)}(\mathbf{c}_i) \right),
\end{equation}
where $\text{EMA}^{(b-1)}(\mathbf{c}_i)$ is the recursively updated activation frequency of intent $\mathbf{c}_i$ up to batch $b$-1, initialized as $\text{EMA}^{(0)}(\mathbf{c}_i) = 0$, $\text{Freq}^{(b)}(\mathbf{c}_i)$ is the activation frequency of $\mathbf{c}_i$ in the current batch, and $\alpha \in (0,1)$ controls the decay rate.
This mechanism smooths noisy fluctuations and facilitates consistent identification of public intents. 
$\mathcal{C}_{pub}$ offers a stable semantic backbone for intent-based modeling.

\subsection{Intent-based Sequence Learning}
In this step, we integrate the retrieved intents into sequence modeling to enhance recommendations. 
We propose multi-branch attentions including several components: (1) a personal intent branch, which injects user-specific intent signals to support personalization and align with individual semantic preferences; 
(2) a public intent branch, which incorporates globally shared intent patterns to improve generalization across users;
and (3) a self-attention branch, which models temporal dependencies in the interaction sequence. 
To combine the information from these branches, we further introduce an adaptive representation fusion layer to integrate the resulting representations into a unified user representation for downstream recommendation.

\textbf{Multi-Branch Attentions}. 
To model different aspects of user behavior, we introduce attentions among temporal dependencies and two types of intents. 
While conventional attention~\cite{vaswani2017attention} can effectively model temporal dependencies in sequences, it lacks the capacity to incorporate intents. 
To this end, we design a new \textit{intent-aware attention layer}.
Specifically, for the personal-user intent attention, we compute alignment between the sequence representation $\mathbf{H}_u \in \mathbb{R}^{T\times d}$ and the personal intent set $\mathcal{C}_{per}^{(u)} \in \mathbb{R}^{S \times d}$ as follows, 
\begin{equation}
    \mathbf{P}_{per}(u) = \text{softmax} \left( \frac{\mathbf{Q}_{per} \mathbf{K}_{per}^\intercal}{\sqrt{d}} \right) \mathbf{V}_{per},
    \label{intent_atten}
\end{equation}
where $\mathbf{Q}_{per}, \mathbf{K}_{per}, \mathbf{V}_{per} = \mathbf{H}_{u}\mathbf{W}_{Q}, \mathcal{C}_{per}^{(u)}\mathbf{W}_{K}, \mathcal{C}_{per}^{(u)}\mathbf{W}_{V}$. 
The output $\mathbf{P}_{per}(u) \in \mathbb{R}^{T \times d}$ denotes the sequence representation derived from personal intents, enhancing the personalization. 

Then, the public intent attention can be designed in a similar way, by replacing $\mathcal{C}_{per}^{(u)}$ with the public intent set $\mathcal{C}_{pub}$ to obtain $\mathbf{P}_{pub}(u) \in \mathbb{R}^{T \times d}$, which is the sequence representation derived from public intents, enhancing the generalization. 
Finally, by replacing the intent set with $\mathbf{H}_u$ in Equation~\ref{intent_atten}, a normal self-attention layer produces a base representation $\mathbf{P}_g(u) \in \mathbb{R}^{T \times d}$.

\textbf{Adaptive Representation Fusion Layer}.
To integrate the above three types of sequence representations, 
we introduce an adaptive fusion layer to balance personalization, generalization, and sequential consistency.
It dynamically weights each branch based on user context $\mathbf{H}_u =[\mathbf{h}_1,...,\mathbf{h}_T]$ as follows, 
\begin{equation}
\begin{split}
    & \mathbf{P}(u) = [\mathbf{P}_{per}(u),\, \mathbf{P}_{pub}(u),\, \mathbf{P}_g(u)], \\ 
    & w = \mathrm{softmax}\!\left(
    \frac{ \mathbf{H}_u\mathbf{P}(u)^{\top} }{\sqrt{d}}
    \right), \\
    & \mathbf{P}_f(u) = \sum_{i=1}^3 w^{(i)} \circ \mathbf{P}^{(i)}(u),
\end{split}
\end{equation}
where $\mathbf{P}(u)\in\mathbb{R}^{T\times 3\times d}$ denotes the stacked branch representations
and $w\in\mathbb{R}^{T\times 3}$ is the adaptive fusion weight. 
The fused representation $\mathbf{P}_f(u)\in\mathbb{R}^{T\times d}$ is used as the final user sequence representation, which preserves sequential consistency and incorporates intent priors for personalized recommendations.

\subsection{Objective Function}
Given the fused user representation $\mathbf{P}_f(u)$, we compute the probability distribution over all items for the next interaction using a softmax function as follows, 
\begin{equation}
\hat{\mathbf{y}}_u = \mathrm{softmax}\!\left( \mathbf{h}_f^u \cdot \mathbf{G}^\top \right),
\label{pred}
\end{equation}
where $\mathbf{h}_f^u \in \mathbb{R}^d$ denotes the final hidden state of $\mathbf{P}_f(u)$, and $\mathbf{G} \in \mathbb{R}^{|\mathcal{V}| \times d}$ is the item embedding matrix. 
The resulting vector $\hat{\mathbf{y}}_u$ represents the predictive score of all items. 
Then, we minimize the cross-entropy loss between the predicted distribution and the ground-truth next item as follows,
\begin{equation}
\mathcal{L}_{\text{Rec}} = -\log\!\left( \hat{\mathbf{y}}_u^\top \mathbf{y}_u \right) + \lambda_\theta \|\theta\|_2^2,
\label{en_loss}
\end{equation}
where $\mathbf{y}_u$ is a one-hot vector indicating the ground-truth next item, and $\theta$ denotes the set of all trainable parameters in the model. 
The hyperparameter $\lambda_\theta$ controls the strength of $\ell_2$ regularization to mitigate overfitting.

\section{Experiments}
In this section, we evaluate the recommendation performance of the proposed SAERec on four public datasets. We further investigate its robustness, explanation quality, and computational efficiency. Finally, we conduct sensitivity analysis and case studies to provide deeper insights into the behavior of SAERec.

\subsection{Experimental Settings}
\subsubsection{Datasets}
We conduct experiments on four public datasets following the experimental settings in \cite{chen2022intent}. 
Specifically, we use the \textbf{Beauty}, \textbf{Toys}, and \textbf{Sports} subsets from the Amazon shopping dataset~\cite{mcauley2015image}, as well as a business recommendation dataset from \textbf{Yelp} \cite{geng2022recommendation}. 
The detailed statistical characteristics of these datasets are summarized in Table~\ref{tab:dataset_stats}.


\begin{table}[t]
\centering
\caption{Statistics of the datasets.} 
\vspace{-7pt}
\label{tab:dataset_stats}
\setlength{\tabcolsep}{5.5pt} 
\renewcommand{\arraystretch}{1.1} 
\begin{tabular}{l|cccc} 
\toprule 
\textbf{Dataset} & \textbf{\# users} & \textbf{\# items} & \textbf{\# Interactions} & \textbf{\# Reviews} \\
\midrule 
Beauty & 22,363 & 12,101 & 198,502 & 158802 \\
Toys   & 19,412 & 11,924 & 167,597 & 134078 \\
Sports & 35,598 & 18,357 & 296,337 & 237070 \\
Yelp   & 30,431 & 20,033 & 316,354 & 253083 \\
\bottomrule 
\end{tabular}
\end{table}

\begin{table*}[t]
\renewcommand{\arraystretch}{1}
\setlength{\tabcolsep}{2pt}
\centering
\captionsetup[table]{skip=2pt}
\vspace{-2pt}

\begingroup
\setlength{\tabcolsep}{2.6pt}
\renewcommand{\arraystretch}{1.25}
\setlength{\aboverulesep}{0.18ex}
\setlength{\belowrulesep}{0.18ex}
\setlength{\cmidrulekern}{0.27em}

\caption{Performance comparison of SAERec and baselines on four datasets (H = HR, N = NDCG).}
\vspace{-2pt}

\resizebox{\textwidth}{!}{

{\scriptsize
\begin{tabular}{l|c c c c|c c c c|c c c c|c c c c}
\toprule
\multirow{2}{*}{Model} &
\multicolumn{4}{c|}{Beauty} &
\multicolumn{4}{c|}{Toys} &
\multicolumn{4}{c|}{Sports} &
\multicolumn{4}{c}{Yelp} \\
\cmidrule{2-17}
 & H@10 & H@20 & N@10 & N@20
 & H@10 & H@20 & N@10 & N@20
 & H@10 & H@20 & N@10 & N@20
 & H@10 & H@20 & N@10 & N@20 \\
\midrule

BPR
& 0.0296 & 0.0474 & 0.0147 & 0.0192
& 0.0197 & 0.0327 & 0.0101 & 0.0132
& 0.0215 & 0.0369 & 0.0105 & 0.0144
& 0.0208 & 0.0346 & 0.0102 & 0.0143 \\

\midrule

GRU4Rec
& 0.0284 & 0.0478 & 0.0150 & 0.0186
& 0.0184 & 0.0290 & 0.0097 & 0.0123
& 0.0258 & 0.0421 & 0.0142 & 0.0174
& 0.0235 & 0.0371 & 0.0113 & 0.0145 \\

Caser
& 0.0342 & 0.0643 & 0.0226 & 0.0298
& 0.0333 & 0.0542 & 0.0168 & 0.0221
& 0.0261 & 0.0399 & 0.0135 & 0.0178
& 0.0241 & 0.0406 & 0.0120 & 0.0156 \\

SASRec
& 0.0624 & 0.0894 & 0.0342 & 0.0386
& 0.0652 & 0.0957 & 0.0320 & 0.0397
& 0.0333 & 0.0500 & 0.0177 & 0.0218
& 0.0252 & 0.0443 & 0.0129 & 0.0179 \\

\midrule

BERT4Rec
& 0.0601 & 0.0984 & 0.0300 & 0.0391
& 0.0524 & 0.0760 & 0.0309 & 0.0368
& 0.0359 & 0.0604 & 0.0190 & 0.0251
& 0.0338 & 0.0564 & 0.0170 & 0.0223 \\

S3-RecMIP
& 0.0307 & 0.0487 & 0.0153 & 0.0198
& 0.0689 & 0.0940 & 0.0383 & 0.0452
& 0.0205 & 0.0344 & 0.0111 & 0.0146
& 0.0214 & 0.0358 & 0.0108 & 0.0144 \\

CL4SRec
& 0.0642 & 0.0974 & 0.0345 & 0.0428
& 0.0736 & 0.0990 & 0.0339 & 0.0404
& 0.0396 & 0.0557 & 0.0191 & 0.0236
& 0.0375 & 0.0639 & 0.0181 & 0.0239 \\

CoSeRec
& 0.0725 & 0.1034 & 0.0401 & 0.0487
& 0.0755 & 0.1037 & 0.0442 & 0.0513
& 0.0439 & 0.0636 & 0.0244 & 0.0302
& 0.0379 & 0.0652 & 0.0183 & 0.0241 \\

DuoRec
& 0.0851 & 0.1228 & 0.0441 & 0.0536
& 0.0959 & 0.1293 & 0.0490 & 0.0574
& 0.0466 & 0.0696 & 0.0244 & 0.0302
& 0.0383 & 0.0674 & 0.0186 & 0.0243 \\

\midrule

P5
& 0.0658 & 0.0825 & 0.0423 & 0.0465
& 0.0690 & 0.0767 & 0.0533 & 0.0553
& 0.0483 & 0.0571 & 0.0331 & 0.0380
& 0.0406 & 0.0657 & 0.0205 & 0.0273 \\

UniSRec
& 0.0547 & 0.0877 & 0.0262 & 0.0345
& 0.0695 & 0.1024 & 0.0319 & 0.0402
& 0.0146 & 0.0257 & 0.0068 & 0.0096
& 0.0331 & 0.0466 & 0.0204 & 0.0238 \\

MoRec
& 0.0297 & 0.0551 & 0.0146 & 0.0210
& 0.0498 & 0.0765 & 0.0234 & 0.0301
& 0.0154 & 0.0272 & 0.0071 & 0.0100
& 0.0173 & 0.0252 & 0.0097 & 0.0117 \\

LLMInit
& 0.0676 & 0.1033 & 0.0311 & 0.0401
& 0.0729 & 0.1069 & 0.0328 & 0.0414
& 0.0403 & 0.0620 & 0.0184 & 0.0238
& 0.0393 & 0.0542 & 0.0203 & \underline{0.0280} \\

AlphaFuse
& 0.0654 & 0.0982 & 0.0304 & 0.0387
& 0.0688 & 0.1010 & 0.0305 & 0.0386
& 0.0383 & 0.0572 & 0.0177 & 0.0224
& 0.0414 & 0.0550 & 0.0209 & 0.0279 \\

\midrule

DSSRec
& 0.0616 & 0.0894 & 0.0326 & 0.0399
& 0.0671 & 0.0942 & 0.0369 & 0.0437
& 0.0328 & 0.0499 & 0.0178 & 0.0215
& 0.0291 & 0.0469 & 0.0135 & 0.0193 \\

SINE
& 0.0612 & 0.0916 & 0.0329 & 0.0384
& 0.0631 & 0.0927 & 0.0383 & 0.0464
& 0.0389 & 0.0610 & 0.0199 & 0.0255
& 0.0302 & 0.0505 & 0.0159 & 0.0216 \\

ICLRec
& 0.0744 & 0.1058 & 0.0403 & 0.0483
& 0.0834 & 0.1139 & 0.0463 & 0.0557
& 0.0437 & 0.0646 & 0.0238 & 0.0301
& 0.0346 & 0.0580 & 0.0174 & 0.0232 \\

IOCRec
& 0.0774 & 0.1146 & 0.0396 & 0.0490
& 0.0804 & 0.1132 & 0.0381 & 0.0464
& 0.0452 & 0.0684 & 0.0220 & 0.0279
& 0.0352 & 0.0617 & 0.0178 & 0.0237 \\

ICSRec
& \underline{0.0953} & 0.1297 & \underline{0.0576} & 0.0663
& \underline{0.1043} & 0.1352 & \underline{0.0652} & 0.0730
& \underline{0.0560} & 0.0787 & \underline{0.0334} & 0.0392
& 0.0421 & 0.0681 & 0.0204 & 0.0269 \\

IRLLRec
& 0.0918 & \underline{0.1328} & 0.0539 & \underline{0.0682}
& 0.1012 & \underline{0.1411} & 0.0618 & \underline{0.0752}
& 0.0541 & \underline{0.0808} & 0.0321 & \underline{0.0400}
& \underline{0.0427} & \underline{0.0690} & \underline{0.0211} & 0.0278 \\

\midrule

SAERec (ours)
& \textbf{0.1055} & \textbf{0.1427} & \textbf{0.0643} & \textbf{0.0737}
& \textbf{0.1147} & \textbf{0.1529} & \textbf{0.0723} & \textbf{0.0820}
& \textbf{0.0611} & \textbf{0.0871} & \textbf{0.0353} & \textbf{0.0424}
& \textbf{0.0452} & \textbf{0.0726} & \textbf{0.0228} & \textbf{0.0297} \\

Improvement
& 10.70\% & 7.45\% & 11.63\% & 8.06\%
& 9.97\% & 8.36\% & 10.89\% & 9.04\%
& 9.11\% & 7.80\% & 5.69\% & 6.00\%
& 5.85\% & 5.22\% & 8.06\% & 6.07\% \\

\bottomrule
\end{tabular}
}
}
\endgroup

\label{tab:main}
\end{table*}

\begin{figure*}[!t]
    \centering
    \includegraphics[scale=0.26]{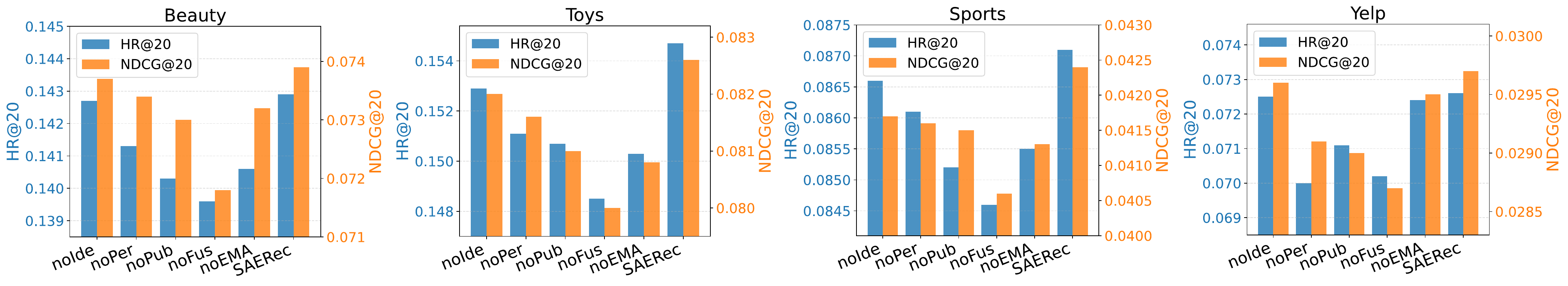} 
    \caption{Ablation studies on four datasets.}
    \label{fig:four-figures}
\end{figure*}

\subsubsection{Evaluation Metrics}
Following \cite{qin2024intent}, we evaluate all models by ranking items over the entire item set without negative sampling. 
We adopt two widely used top-$k$ ranking metrics, Hit Ratio (HR@$k$) and Normalized Discounted Cumulative Gain (NDCG@$k$), with $k \in \{10, 20\}$. 
This evaluation protocol is commonly employed in intent-based recommendation studies \cite{tan2021sparse, li2023multi, qin2024intent}, ensuring fair and consistent comparisons with existing methods.

\subsubsection{Baselines} 
We compare SAERec with representative methods from the following categories. 
(1) Non-sequential models. 
\textit{BPR} \cite{rendle2012bpr} is a classical matrix factorization method optimized with Bayesian Personalized Ranking loss.
(2) Standard sequential models. 
\textit{GRU4Rec} \cite{hidasi2015session} is an RNN-based approach, 
\textit{Caser} \cite{tang2018personalized} adopts convolutional architectures, 
and \textit{SASRec} \cite{kang2018self} is a state-of-the-art Transformer-based sequential recommender. 
(3) Sequential models with self-supervised learning (SSL).
\textit{BERT4Rec} \cite{sun2019bert4rec} applies masked item prediction to capture bidirectional dependencies.
\textit{S$^3$-Rec} \cite{zhou2020s3}, \textit{CL4SRec} \cite{xie2022contrastive}, \textit{CoSeRec} \cite{liu2021contrastive}, and \textit{DuoRec} \cite{qiu2022contrastive} incorporate various contrastive or augmentation-based SSL objectives to enhance sequential representation learning. 
(4) Sequential models with large language models (LLMs). 
\textit{P5} \cite{geng2022recommendation} formulates recommendation as a prompt-based generation task.
\textit{UniSRec} \cite{hou2022towards} introduces a mixture-of-experts adaptor,
\textit{MoRec} \cite{yuan2023go} leverages item modality features,
\textit{LLMInit} \cite{hu2024enhancing} initializes item embeddings using language representations,
and \textit{AlphaFuse} \cite{hu2025alphafuse} decomposes language embeddings to better integrate semantic and collaborative signals. 
(5) Sequential models with latent intents. 
\textit{DSSRec} \cite{ma2020disentangled} disentangles user intentions via seq2seq modeling.
\textit{SINE} \cite{tan2021sparse} captures multiple user interests through adaptive aggregation.
\textit{ICLRec} \cite{chen2022intent}, \textit{IOCRec} \cite{li2023multi}, and \textit{ICSRec} \cite{qin2024intent} explicitly model latent intents and incorporate intent-level contrastive learning for sequential recommendation. 
\textit{IRLLRec} \cite{wang2025intent} further leverages large language models to learn textual intent representations and aligns them with interaction-based intents for recommendation.

\subsubsection{Implementation Details} 
We adopt a unified implementation setting across all datasets and experiments. 
For \textbf{sparse autoencoder (SAE) training}, we employ the Top-$K$ sparsity formulation and fix the target sparsity level to $K = 10$ for all datasets, following common practice in prior SAE studies \cite{bricken2023towards,lieberum2024gemma}. 
The capacity of the SAE is determined by a scaling law \cite{gao2024scaling} that relates the number of SAE features $M$ to the number of training tokens $P$ in the review corpus \cite{bricken2023towards,lieberum2024gemma}.
Specifically, following empirical observations in prior work, we set $M = P^{\gamma}$, where $\gamma \approx 0.60$ for GPT2-small and $\gamma \approx 0.65$ for GPT-4 \cite{gao2024scaling}.
This procedure yields $M = 2^{14}$ for Beauty and Toys, and $M = 2^{15}$ for Sports and Yelp, without dataset-specific tuning. 
The size of the extracted word set is fixed to $Z = 10$ following previous work \cite{wu2025self}. And we use Mistral-7B \cite{jiang2023mistral} as the language model for intent filtering.
For \textbf{recommender training}, we set the number of retrieved intents to $S = 5$.
We employ a lightweight self-attention architecture with 2 self-attention blocks and 2 attention heads.
The embedding dimension is set to $d = 64$, and the maximum sequence length is fixed to 50.
All architectural hyperparameters are kept identical across datasets and methods to ensure fair comparison. 
To prevent data leakage, the review corpus is constructed using only reviews associated with users and items appearing in the training set.
The model is optimized using the Adam \cite{diederik2014adam} optimizer with a learning rate of $1\times10^{-3}$ and a weight decay coefficient $\lambda_{\theta} = 1\times10^{-4}$.
All experiments are conducted on an NVIDIA RTX A6000 GPU.

\subsection{Comparison With Baselines} 

The evaluation results of SAERec and the SOTA baselines on four datasets are presented in Table 2. 
The best and the second-best results are highlighted in bold and underline. 
We can see that SAERec consistently achieves the best performance in all cases. 

Compared with the strongest competitor, 
SAERec yields substantial gains. For HR@10, it improves by 10.70\%, 9.97\%, 9.11\% and 5.85\% on Beauty, Toys, Sports and Yelp, respectively. 
For NDCG@10, the improvements range from 5.69\% to 11.63\%, and for HR@20 and NDCG@20, it achieves 5.22\% to 9.04\% gains across datasets. 
These improvements are especially notable on sparse datasets like Beauty and Toys, where user intents are diverse but behavioral signals are insufficient, demonstrating the benefit of intent priors modeling that provides stronger guidance for ranking items. 
Compared with contrastive learning-based models (e.g., CL4SRec and DuoRec), which rely on random sequence augmentations, SAERec leverages interpretable intent vectors derived from review semantics. 
These intent priors serve as informative auxiliary signals to guide downstream modeling, which improves generalization under data sparsity and behavior variability.
Compared to LLM-based methods (e.g., MoRec, LLMInit, and AlphaFuse) that improve recommendation by treating text as side information to enhance item representations, SAERec decomposes review semantics into sparse and interpretable semantic factors and explicitly retrieves relevant intent priors during inference. 
This design enables SAERec to maintain fine-grained and reusable semantic guidance under sparse and noisy environments, resulting in consistently better recommendation performance.
Recent intent-aware methods (e.g., ICLRec, IOCRec, ICSRec, and IRLLRec) also incorporate user intents, but they mainly model intents as latent or dense semantic representations inferred from interaction sequences or multimodal alignment.
Such intent representations are heavily affected by sparse or noisy user behaviors and lack explicit semantic interpretability.
In contrast, SAERec derives sparse intent priors directly from large-scale review corpora, providing more comprehensive, fine-grained, and interpretable semantic signals for downstream recommendation.

\begin{table}[t]
\centering
\caption{Comparison with Clustering-based Intent Construction.} 
\vspace{-2pt}
\setlength{\tabcolsep}{8pt}
\renewcommand{\arraystretch}{1.0}
\begin{tabular}{l|l|cc}
\toprule
\textbf{Data} & \textbf{Metric} & \textbf{KMeans-Intent} & \textbf{SAERec}   \\
\midrule
\multirow{2}{*}{Beauty} & H@20    & 0.1349 & 0.1427 \\
                        & N@20  & 0.0671 & 0.0737 \\
\midrule
\multirow{2}{*}{Sports} & H@20    & 0.0818 & 0.0871 \\
                        & N@20  & 0.0398 & 0.0424 \\

\bottomrule
\end{tabular}

\label{tab:replace}
\end{table}


\subsection{Ablation Study} 

We further conduct an ablation study to evaluate the contribution of each component in SAERec to the performance gain. 
We deactivate different components and form the following variants: 
\textbf{noIde} removes the semantic intent filtering stage and uses all factors learned from reviews as intent priors; 
\textbf{noPer} removes the personal intent guidance; 
\textbf{noPub} removes the public intent guidance; 
\textbf{noFus} disables the adaptive fusion and replaces it with simple summation; 
\textbf{noEMA} removes the EMA-based weighting in intent retrieval.

The results of the ablation study are shown in Figure \ref{fig:four-figures}. 
All components are useful for SAERec as removing any one of them results in a performance decline. 
Among these, 
we can see that noIde exhibits consistent drops across datasets, underscoring the importance of LLM-based intent identification. 
Although relevant intents can still be retrieved, the inclusion of noisy or less-informative vectors hinders retrieval precision and weakens recommendation. 
It highlights the benefit of intent identification in improving semantic quality and enhancing interpretability.
And we observe that noPer leads to drops in performance, indicating the importance of personal intent, where personalized signals effectively capture fine-grained user motivations critical for sequence modeling.  
Similarly, noPub also brings noticeable drops, highlighting the role of shared generic intents in enhancing generalization. 
Disabling the adaptive fusion layer yields further decreases, showing that simply summing the representations is insufficient, and adaptive fusion is crucial for dynamically balancing personalization and generalization based on user context.  
The noEMA shows marginally lower performance, confirming that EMA-based weighting helps filter generic intents and stabilize retrieval across batches. 
Overall, the full SAERec consistently achieves the best results, validating the effectiveness of our intent construction and integration.

\subsection{Intent Construction Analysis}

To further evaluate whether SAE is necessary for intent construction, we replace the SAE-based sparse decomposition with a simpler clustering-based strategy.
Specifically, we apply KMeans clustering over the LLM-derived review embeddings, treat the cluster centroids as intent candidates, and keep the remaining retrieval and recommendation pipeline unchanged. 
The results are shown in Table~\ref{tab:replace}. 
We observe that KMeans-Intent consistently underperforms SAERec on both Beauty and Sports. This suggests that simple clustering is insufficient for constructing fine-grained intent priors. 
We attribute this difference to the distinct modeling mechanisms. KMeans assigns each review embedding to a single cluster centroid, which mainly captures coarse semantic grouping. 
In contrast, SAE decomposes each review into sparse and overcomplete semantic factors, allowing multiple fine-grained intent signals to coexist within the same review.
This enables SAERec to construct more specialized and reusable intent priors for downstream recommendation. 
Moreover, the sparse decomposition produced by SAE naturally supports semantic interpretation through aligned keywords, making the resulting intent priors more human-understandable than clustering-based semantic grouping. As a result, SAERec achieves both better recommendation performance and improved interpretability. 
\begin{table*}[t]
\centering
\caption{Samples of learned intent priors with top keywords and semantic themes.}
\vspace{-7pt}
\renewcommand{\arraystretch}{1.1}
{\footnotesize
\begin{tabularx}{\textwidth}{l|L{0.60\textwidth}|L{0.30\textwidth}}
\toprule
\textbf{Dataset} & \centering \textbf{Top-10 keywords} & \textbf{Automatically Summarized Intent} \\
\midrule

\multirow{5}{*}[-3.0ex]{Beauty}
& almay, reviva, oxybenzone, luminary, shiseido, benzyl, weightless, lifeless, avobenzone, rosy.
& Skincare and sunscreen-related. \\
\cmidrule(lr){2-3}

& favorable, oriental, lancome, turquoise, unscented, muted, parfum, harmony, fragrant, testing.
& Fragrance preferences and exploration. \\
\cmidrule(lr){2-3}

& headband, badger, plump, mechanical, phenoxyethanol, matrix, clipper, haircolor, hairdresser, extent.
& Hair care and styling. \\
\cmidrule(lr){2-3}

& smoothness, cuticle, smoothing, vaseline, reapply, itching, gelish, drugstore, holding, repurchase.
& Nail care and maintenance. \\
\cmidrule(lr){2-3}

& holding, absorbs, wrinkles, moisturizer, breakage, thinning, staying, beautiful, moist, purchase.
& Anti-aging and skincare. \\

\specialrule{0.8pt}{2.2pt}{2.2pt}

\multirow{5}{*}[-3.0ex]{Sports}
& cyclist, cycling, interval, control, calculate, winner, expectation, remarkably, surprisingly, plenty.
& Cycling performance optimization and training. \\
\cmidrule(lr){2-3}

& travelling, hiking, trekking, walking, hike, stepping, sleeping, digging, slipping, travel.
& Outdoor physical activities, hiking and trekking. \\
\cmidrule(lr){2-3}

& swim, endurance, interval, lessons, efficiency, pace, skill, experience, execution, dealing.
& Swimming skills and endurance improvement. \\
\cmidrule(lr){2-3}

& fitbit, athlete, wore, strapped, resolution, voltage, led, leakage, gauge, extension.
& Fitness tracking and measurement tools. \\
\cmidrule(lr){2-3}

& backpack, camping, pack, kit, rack, yards, keeping, cool, expensive, pump.
& Camping and backpacking equipment. \\

\bottomrule
\end{tabularx}
}
\label{tab:intent_priors_examples}
\end{table*}

\begin{table}[t]
\centering
\caption{Robustness of SAERec under different ratio of data sparity.} 
\vspace{-2pt}
\setlength{\tabcolsep}{8pt}
\renewcommand{\arraystretch}{1.0}
\begin{tabular}{l|l|cccc}
\toprule
\textbf{Data} & \textbf{Metric} & \textbf{0\%} & \textbf{10\%} & \textbf{20\%} & \textbf{30\%}  \\
\midrule
\multirow{2}{*}{Beauty} & H@20    & 0.1427 & 0.1413 & 0.1431 & 0.1441 \\
                        & N@20  & 0.0737 & 0.0729 & 0.0735 & 0.0740\\
\midrule
\multirow{2}{*}{Sports} & H@20    & 0.0871 & 0.0876 & 0.0877 & 0.0882\\
                        & N@20  & 0.0424 & 0.0425 & 0.0425  & 0.0426\\

\bottomrule
\end{tabular}

\label{tab:robust}
\end{table}

\subsection{Robustness Analysis}
We evaluate the robustness of SAERec on the sparse data by adding the perturbation to the raw datasets through the deletion operations. Specifically, we set a ratio as $r$\% and randomly select $r$\% percentage of the reviews to remove, simulating the scenario of sparse data. 
Based on Beauty and Sports datasets, the $r$ is set to $\{0, 10, 20, 30\}$, and the results are shown in Table~\ref{tab:robust}. 
We can see that SAERec remains remarkably stable across all cases.  
On Beauty, H@20 fluctuates within 2.0\% and N@20 within 1.6\%. On Sports, the ranges are within 0.7\% for H@20 and 0.3\% for N@20. It indicates that SAERec’s performance is stable to random review missingness.   
We notice that even with 30\% review removal, the performance of SAERec is effectively unchanged and occasionally improves slightly (such improvements may stem from incidental noise pruning with random deletion, as reviews naturally contain much noisy signals). 
We attribute this resilience to learning a fine-grained interpretable intent set from LLM embeddings: reviews are distilled into intent priors rather than used as raw auxiliary text. And the personal, public intents retrieved and fusion further suppress noisy or redundant signals. 
These results demonstrate that SAERec is reliable when review coverage is limited.

\subsection{Explanation Analysis}
In this section, we analyze the constructed intent set $\mathcal{C}^*$ to examine its semantic interpretability. 
Table~\ref{tab:intent_priors_examples} presents some examples learned from the Beauty and Sports datasets, where each intent prior is characterized by a set of keywords and captures a coherent and human-understandable semantic theme. 
For the Beauty dataset, the learned intent priors reflect fine-grained, product-oriented user preferences. 
Representative examples include intents related to skincare and sunscreen usage, hair care and styling routines, as well as fragrance exploration and nail care.
These themes align closely with common concerns expressed in user reviews and purchasing behaviors, indicating that the learned priors capture meaningful aspects of user intent.  
For the Sports dataset, the learned intent priors similarly correspond to clear, activity-centric user preferences. 
Examples include cycling performance and interval training, outdoor hiking and trekking activities,
swimming skills and endurance improvement, fitness tracking and measurement tools, 
and camping or backpacking equipment. 
These intents reflect users’ diverse goals across training, outdoor recreation, and performance monitoring. 
This qualitative analysis confirms that the constructed intent set $\mathcal{C}^*$ exhibits a meaningful semantic structure. 

To further examine whether the learned intents are actionable in recommendations, we conduct \textbf{counterfactual experiments} on the Beauty dataset.
Specifically, we up-weight intents aligned with a target category during inference and compare the recommendation distributions before and after the manipulation. 
We observe that boosting \emph{Skin Care}-aligned intents increases the number of recommended Skin Care items from 6{,}819 to 7{,}011, and boosting \emph{Tools \& Accessories}-aligned intents increases the number of recommended Tools \& Accessories items from 3{,}151 to 3{,}474.
These shifts indicate that the learned intent priors are not only human-readable but also actionable, as intervening on them can steer category-level recommendation distributions without retraining the model.

\begin{figure}[!t]
  \centering

\begin{subfigure}{\linewidth}
    \centering
    \includegraphics[scale=0.333]{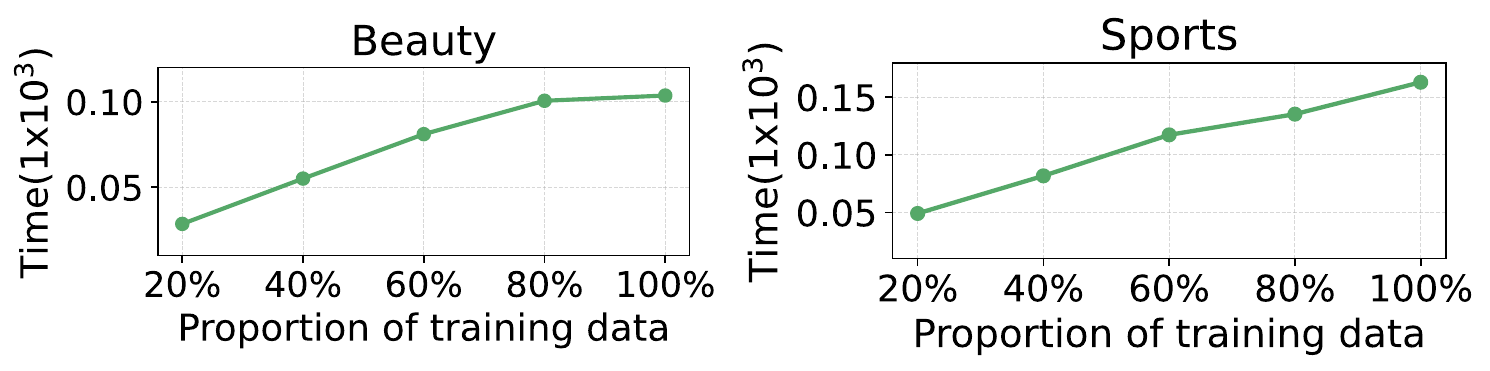}
    \caption{Training time on SAERec.}
    \label{saerec-time}
  \end{subfigure}

  \vspace{-1pt}

\begin{subfigure}{\linewidth}
    \centering
    \includegraphics[scale=0.33]{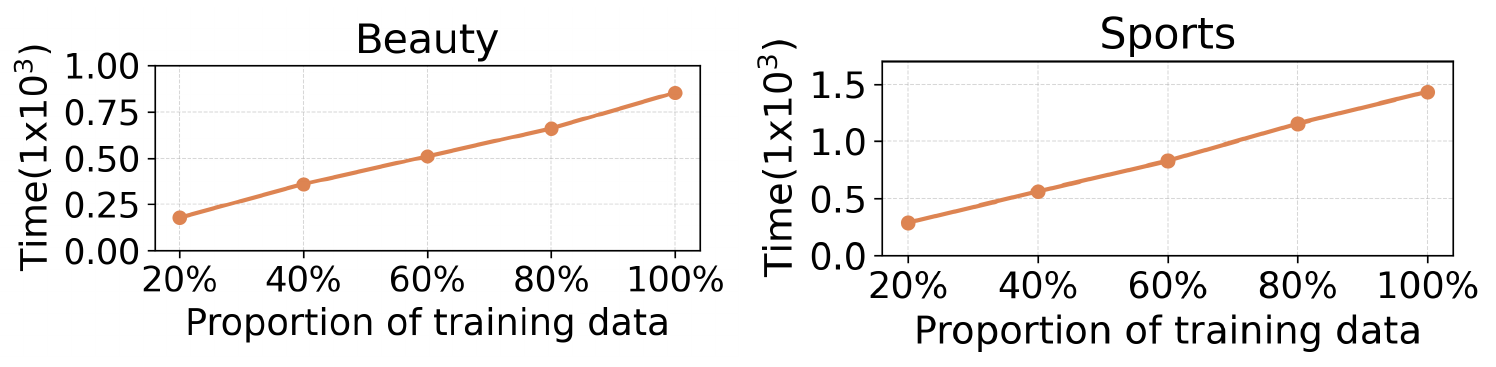}
      \caption{Training time on SAE.}
    \label{fig:overall:b}
  \end{subfigure}
  \caption{Efficiency analysis.} 
  \label{fig:Training-efficiency}
\end{figure}

\subsection{Model Efficiency Analysis}
To investigate the time efficiency of SAERec, we train the
model on Beauty and Sports by varying the proportion of training data
in \{20\%, 40\%, 60\%, 80\%, 100\%\} and report the average training
time for each epoch. 
The average training time curve for each epoch with the increasing amount of training data is shown in Figure \ref{fig:Training-efficiency} (a). 
One can see that the training time increases nearly linearly with the proportion of training data on both Beauty and Sports datasets. 
This confirms that SAERec achieves a desirable balance between effectiveness and efficiency, making it practical for large-scale real-world recommendation scenarios where both accuracy and training efficiency are critical. 

To quantify the cost of constructing the intent set, we train the SAE under different review retention ratios \{20\%, 40\%, 60\%, 80\%, 100\%\}. 
Figure \ref{fig:Training-efficiency} (b) reports the average time per epoch. We find that SAE training time also scales approximately linearly with the amount of review data. In addition, the memory cost is small in practice. We find the peak GPU memory during SAE training stays below 3 GB across all retention ratios and datasets, which fits comfortably on a single modern GPU. 
This is consistent with the TopK SAE ($M \gg K$), so per-step compute and activation storage scale with $O(D \times K)$ rather than $O(D \times M)$. 
Overall, these results indicate that SAERec introduces only moderate and scalable overhead for intent construction, where the intent priors are constructed \textbf{only once and can be reused}, preserving efficient model training and practical feasibility for large-scale recommendation scenarios.

\begin{table}[t]
\centering
\caption{Performance of SAERec with different language models.} 
\vspace{-2pt}
\setlength{\tabcolsep}{6.3pt}
\renewcommand{\arraystretch}{1.0}
\begin{tabular}{l|l|cccc}
\toprule
\textbf{Data} & \textbf{Metric} & \textbf{Mistral} & \textbf{LLama} & \textbf{Qwen} & \textbf{Gemma}  \\
\midrule
\multirow{2}{*}{Beauty} & H@20    & 0.1427 & 0.1416 & 0.1422 & 0.1420 \\
                        & N@20  & 0.0737 & 0.0724 & 0.0732 & 0.0729\\
\midrule
\multirow{2}{*}{Sports} & H@20    & 0.0871 & 0.0885 & 0.0887 & 0.0869 \\
                        & N@20  & 0.0424 & 0.0426 & 0.0431 & 0.0422 \\

\bottomrule
\end{tabular}

\label{tab:llm}
\end{table}

\subsection{Sensitivity Analysis}

We first examine the sensitivity of SAERec with respect to the choice of language models used for intent filtering. 
Specifically, we consider four representative LLMs, including Mistral-7B, LLaMA-7B \cite{touvron2023llama}, Qwen-7B \cite{bai2023qwen}, and Gemma-7B-IT \cite{team2024gemma}. 
The corresponding results are reported in Table~\ref{tab:llm}. 
The observed performance differences across language models are consistently small across both datasets. 
On the Beauty dataset, the maximum relative variation across language models is 0.78\% for HR@20 and 1.80\% for NDCG@20.
On the Sports dataset, the corresponding variations are 2.07\% for HR@20 and 2.13\% for NDCG@20. 
These results indicate that SAERec exhibits stable performance across different language model. 

\begin{figure}[t]
    \centering
    \includegraphics[scale=0.237]
    {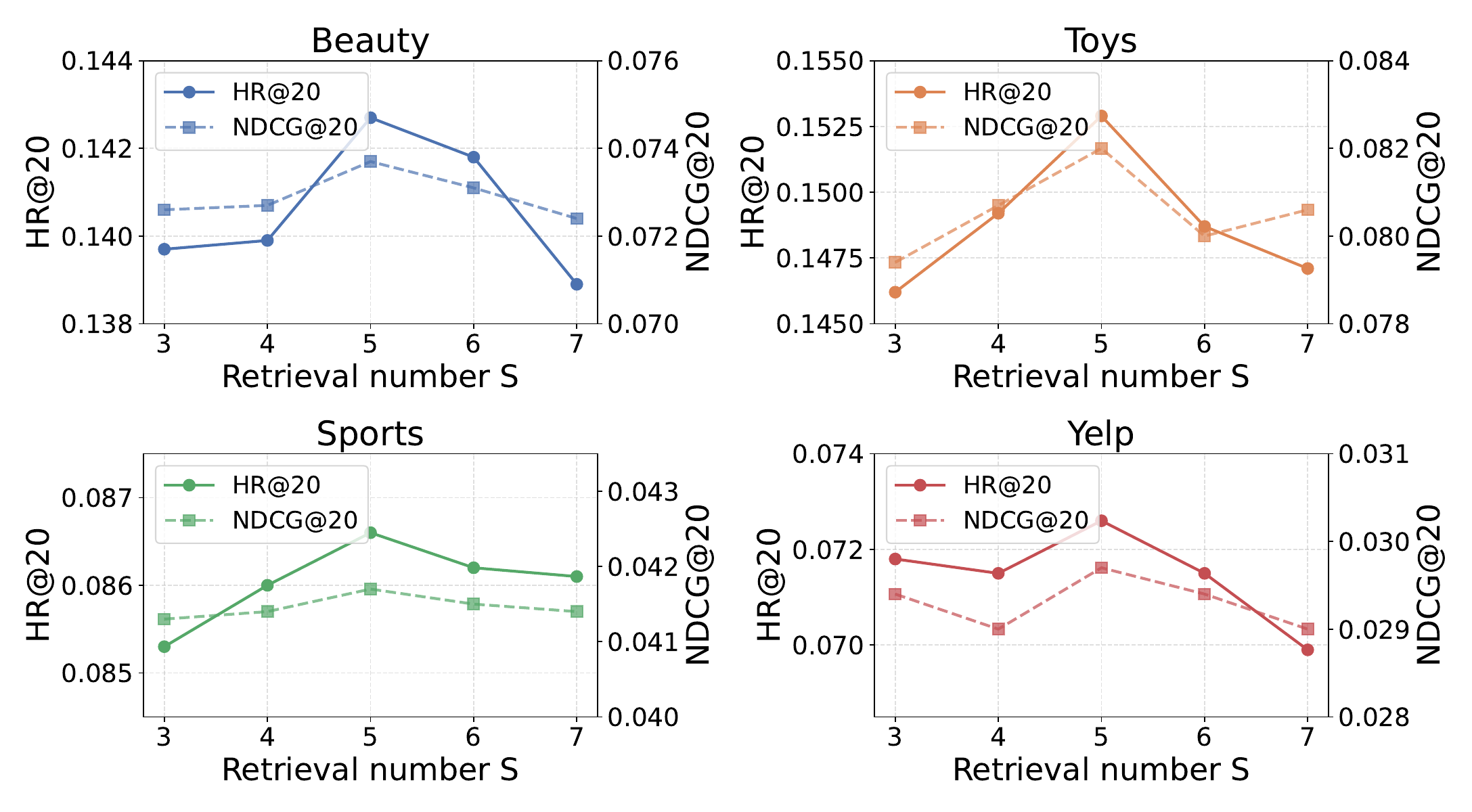}
    \caption{Effect of different retrieval numbers S.}
    \label{fig:sensitivity_topk}
\end{figure}

We further investigate the impact of retrieval hyperparameter $S$, which controls the number of retrieved intents to guide recommendation. 
The results are shown in Figure \ref{fig:sensitivity_topk}.
For $S$, by varying it from 3 to 7, we found the optimal setting for achieving the best performance was at $S$=5. 
Across datasets, we observe improvements when increasing 
$S$ from 3 to 5,
as retrieving more intents enriches the model's semantic guidance, allowing the recommender to incorporate diverse, relevant user intents to improve ranking accuracy.
Performance declined when $S$ increases beyond 5, it is possibly due to the inclusion of less relevant or noisy intents when $S$ becomes large, which may dilute the effectiveness of the most informative intents.


\subsection{Case Study}
\begin{figure}[!t]
    \centering
    \includegraphics[scale=0.163]{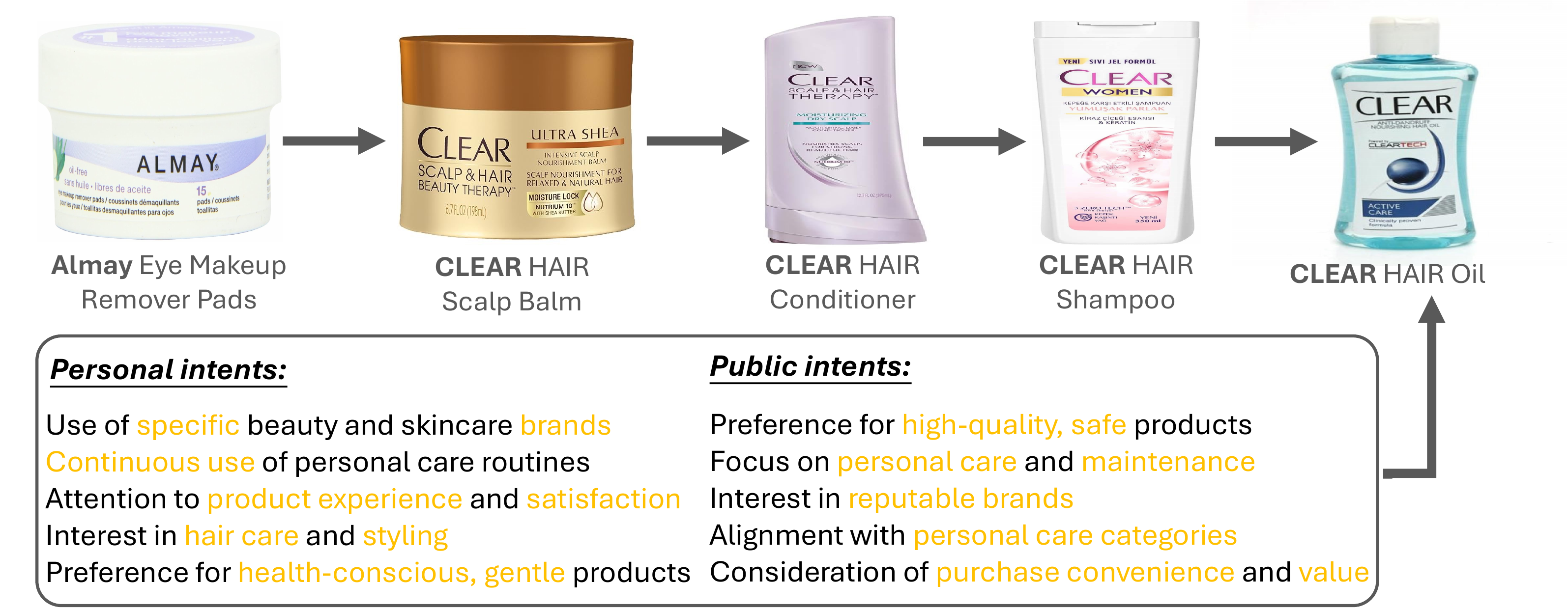}
    \caption{Recommendation of $u_{580}$.}
    \label{case2}
\end{figure}

We conduct a case study to qualitatively evaluate the effectiveness of SAERec.
Figure~\ref{case2} illustrates how SAERec performs intent priors-guided modeling in practice.
One can see that SAERec retrieves both personal intents (e.g., "brand trust", "haircare routines") and public intents (e.g., "quality", "safety") from the learned intent priors.
These retrieved intents are highly consistent with the user’s historical interactions, indicating that the model effectively captures the user’s underlying motivations while incorporating commonly shared concerns across users. 
Guided by these intents, SAERec recommends CLEAR Hair Oil, which is consistent with the user’s previous purchases from the same brand and well aligned with their ongoing haircare routines.
Specifically, this recommendation satisfies personal intents by reinforcing the user’s trust in the CLEAR brand and supporting their habitual haircare preferences. 
At the same time, it satisfies public intents by providing a high-quality and reliable product that conforms to general care standards.
These results demonstrate that SAERec effectively leverages intent priors to deliver interpretable recommendations that align with both personal and public intents.

\section{Related Work}
\textbf{Sequential Recommendation.} Early sequential recommenders modeled next-item prediction from interaction histories using RNNs (e.g., GRU4Rec) \cite{hidasi2015session}. Self-attention then became dominant with SASRec, capturing long-/short-range dependencies \cite{kang2018self}; BERT4Rec further exploited bidirectional transformers for sequence encoding \cite{sun2019bert4rec}. Recently, LLM-based approaches inject richer semantics from text to improve recommendations. 
Some researches regard recommendation tasks as language generation tasks such as P5 \cite{geng2022recommendation}, while others fuse text or multimodal embeddings into collaborative or sequential architectures (UniSRec \cite{hou2022towards}, MoRec \cite{yuan2023go}), or leverage text embeddings for initialization or decomposition alongside trainable item representations (LLMInit \cite{hu2024enhancing}, AlphaFuse \cite{hu2025alphafuse}). These methods improve semantic capacity but usually treat text as side information to enhance recommendations. 


\textbf{Intent-based Recommendation.} Understanding users' underlying intents has emerged as a critical direction to enhance recommendation \cite{jannach2024survey}. 
Most researchers model intent as latent variables based on user sequences. 
ICLRec constructs user intents via sequence clustering and integrates them into the recommendation process through a contrastive self-supervised learning loss \cite{chen2022intent}. 
IOCRec further improve it, they refine the contrastive framework by introducing global and local intent modules, which are designed to better capture intent semantics during learning \cite{li2023multi}. 
ICSRec aligns cross-subsequences with the same target to learn fine- and coarse-grain intent representations via contrastive learning \cite{qin2024intent}. 
\cite{sun2024large} adopts LLM-based multi-round prompts to infer user-specific intents from item titles and based on intents to re-rank candidate items for recommendation. 


\textbf{Sparse Autoencoders in Recommendation.} 
Sparse autoencoders (SAEs) have recently been explored as a tool for interpreting internal representations of recommendation models \cite{wang2024interpret, klenitskiy2025sparse}.
For example, recent work applies SAEs to post-hoc analyze hidden states of trained recommenders and summarizes activated features to provide descriptive explanations of model behavior.
These approaches primarily focus on understanding or labeling latent features after model training, without feeding the learned representations back into the recommendation process \cite{wang2024interpret}. 
In contrast, our work employs SAEs as a constructive component to explicitly build a set of fine-grained, interpretable intent priors from review corpus. 
The learned intent representations are not only human-interpretable at the word level, but are also directly integrated into the recommendation pipeline to guide prediction.
This distinction positions our method beyond post-hoc interpretation, toward intent-based modeling that actively influences recommendation.


\section{Conclusion}
This paper introduces the Sparse Autoencoder for intent-based recommendation
(SAERec), a novel recommender that automatically constructs a comprehensive set of fine-grained interpretable intents from user reviews. 
By leveraging SAE and LLM, the model decouples intent discovery from user-item interactions and enables unsupervised extraction of intent representations. 
These intents are then selectively retrieved and serve as priors to guide recommendation by being injected to sequence modeling, 
effectively balancing personalization, generalization and sequential consistency. 
Extensive evaluations demonstrate the effectiveness of SAERec across diverse datasets, which achieve consistent improvements in recommendation accuracy and explainability.

\appendix

\section{Details of Intent Filtering} 
\label{ap_prompt}

\subsection{Interpreting Learned Vectors} 
Each disentangled vector $\mathbf{c}_m$ learned by the sparse autoencoder is associated with a set of informative review words $\mathcal{I}_m$, which serves as a textual explanation of its semantic meaning. 
Building on prior work in LLM-as-a-judge \cite{bills2023language, chaudhary2024evaluating, gao2024scaling, lieberum2024gemma}, we interpret the learned vectors based on their textual explanation $\{\mathcal{I}_m\}_{m=1}^{M}$. 
To summarize the underlying patterns of $\mathcal{I}_m$, we employ Mistral-7B as our machine annotator, with a temperature of 0 for deterministic decoding. 
To enhance the reliability of this process, the annotation is guided by clear
instructions that emphasize semantic coherence. 
Following previous work \cite{bills2023language}, our machine annotator has the option to respond with ``Cannot Tell'' if it detects no meaningful patterns among the $\mathcal{I}_m$. 
This interpretation step is performed independently for each vector and does not alter the learned vectors. 

\subsection{Filtering Intent Vectors} 
Not all vectors learned by the SAE correspond to meaningful user
intents.
For recommendation, we define intent vectors as those that exhibit a clear semantic association
with user intents. 
Following prior work, we adopt the annotation framework from \cite{lieberum2024gemma}, where the machine annotator classifies correlations into four levels: "Yes", "Probably", "Maybe", and "No".
Vectors are considered meaningful intent if they are assigned a relevance level as "Yes", while the remaining vectors are discarded.  
This filtering step is performed once, and the resulting
intent set $\mathcal{C}^*$ remains fixed during subsequent model training and inference.

\subsection{Prompt Templates}

To improve reproducibility, we provide the prompts used for semantic interpretation and intent filtering. 
The same prompt templates are applied across all datasets, with only the dataset name replaced accordingly.
\vspace{7pt}

\noindent
\begin{tcolorbox}[
width=\linewidth,
breakable,
colback=gray!5!white,
colframe=black!60,
title=Semantic Interpretation Prompt,
fonttitle=\bfseries,
left=3pt,right=3pt,top=3pt,bottom=3pt,
boxsep=2pt
]
\footnotesize
\textbf{System instruction:}
You are analyzing latent semantic factors learned from user reviews in a recommendation dataset.
Each latent vector captures a particular concept, topic, user preference, or behavioral pattern reflected in user reviews.
Look at the following words associated with a latent vector and summarize the underlying semantic concept.
Pay more attention to the words appearing earlier in the list, as they are more strongly correlated with the latent semantic factor.
If the words do not form a coherent semantic pattern, answer ``Cannot Tell''.

\vspace{2pt}
\textbf{Target input:}
\texttt{Words: \{word list\}}

\vspace{2pt}
\textbf{Target output:}
\texttt{\{LLM-generated semantic summary or ``Cannot Tell''\}}
\end{tcolorbox}

\vspace{3pt}

\noindent
\begin{tcolorbox}[
width=\linewidth,
breakable,
colback=gray!5!white,
colframe=black!60,
title=Intent Filtering Prompt,
fonttitle=\bfseries,
left=3pt,right=3pt,top=3pt,bottom=3pt,
boxsep=2pt
]
\footnotesize
\textbf{System instruction:}
You are given a semantic description of a latent feature learned from user reviews in a recommendation dataset.
Please determine whether this feature represents a meaningful user intent relevant to recommendation.

A meaningful intent should reflect coherent user preferences, product characteristics, usage scenarios, purchasing motivations, or decision-related concerns.

Please classify the feature into one of the following categories:
\textbf{Yes}: clearly represents a meaningful recommendation intent.
\textbf{Probably}: likely related to recommendation intent but not sufficiently clear.
\textbf{Maybe}: weakly related or ambiguous.
\textbf{No}: unrelated, noisy, or not meaningful for recommendation.

\vspace{2pt}
\textbf{Target input:}
\texttt{Description: \{semantic summary\}}

\vspace{2pt}
\textbf{Target output:}
\texttt{\{Label: Yes/Probably/Maybe/No\}}
\end{tcolorbox}

\bibliographystyle{ACM-Reference-Format}
\bibliography{saerec-base2}


\end{document}